\documentclass[preprint,5p,twocolumn]{elsarticle}

\usepackage{mathrsfs,amsmath,amsthm,amssymb,braket,color,verbatim,adjustbox,multirow,enumerate,textcomp,gensymb,subfigure,graphicx,diagbox,pifont,tabulary,booktabs,stackengine,epstopdf,dcolumn,makecell,listings,bbm,mathtools,autobreak,slashed,float,array,colordvi,xfrac,orcidlink}
\usepackage{lineno}
\modulolinenumbers[5]
\usepackage{hyperref}
\hypersetup{colorlinks,bookmarksopen,bookmarksnumbered,citecolor=[rgb]{0,0.35,0.5},linkcolor=[rgb]{0,0.35,0.5},urlcolor=[rgb]{0,0.35,0.5},pdfstartview=FitH,linktocpage}

\definecolor{eprintLinks}{rgb}{0,0.35,0.5}
\definecolor{journalLinks}{rgb}{0,0.35,0.5}
\newcommand{\MYhref}[3][blueLinks]{\href{#2}{\color{#1}{#3}}}

\def\g{\gamma}

\def\Ms{M_{s}}
\def\Md{M_{D}}
\def\gs{g_{s}}
\def\O{{\cal O}}

\def\d{\delta}

\def\Elv{E_{\textrm{LV}}}
\def\Ep{E_{\textrm{Pl}}}

\usepackage{etoolbox}
\usepackage{expl3}
\ExplSyntaxOn
\cs_new_eq:NN \Repeat \prg_replicate:nn
\ExplSyntaxOff
\makeatletter
\ifdoubleblind\else
\patchcmd{\tnotetext}{\ifcase\c@tnote\or$\star$\or$\star\star$\fi}{\Repeat{\the\c@tnote}{\tnotemarksymbol}}{}{\@latex@error{Failed to patch \string\tnotetext}}
\patchcmd{\tnotemark}{\expandafter\ifcase\elsRef{\mytmark}\or$^{\star}$\or$^{,\star\star}$\fi}{\textsuperscript{\expandafter\ifnum\elsRef{\mytmark}>1,\fi\expandafter\Repeat\expandafter{\elsRef{\mytmark}}{\tnotemarksymbol}}}{}{\@latex@error{Failed to patch \string\tnotemark}}
\fi
\makeatother
\newcommand{\tnotemarksymbol}{\ding{73}}

\journal{PLB, published as Phys. Lett. B 879 (2026) 140676, \href{https://doi.org/10.1016/j.physletb.2026.140676}{doi:10.1016/j.physletb.2026.140676}}

\begin{document}
\hypersetup{bookmarksopen,bookmarksnumbered,citecolor=[rgb]{0,0.35,0.5},linkcolor=[rgb]{0,0.35,0.5},urlcolor=[rgb]{0,0.35,0.5},linktocpage}
\biboptions{numbers,sort&compress}

\begin{frontmatter}

\title{Remarks on atmospheric effect of D-foam in light of muon puzzle\tnoteref{t1}}
\tnotetext[t1]{This letter is an \emph{addendum} to `Shower formation in the presence of a string-inspired foam in space-time', in \href{https://doi.org/10.1016/j.physletb.2025.139823}{Phys. Lett. B 869 (2025) 139823}.}

\author[pku]{Chengyi Li\,\orcidlink{0000-0002-7222-1094}\corref{em}}
\cortext[em]{Corresponding author \ead{lichengyi@pku.edu.cn}}

\address[pku]{School of Physics, Peking University, Beijing 100871, China}

\begin{abstract}
In our recent paper~\cite{Li:2025wqc}, we used a stringy model for quantum space-time foam to suggest that the so-induced subluminal Lorentz violation~(LV) for photons would not lead to experimentally unacceptable changes in the developments of particle showers initiated by cosmic $\g$-rays in the Earth's atmosphere, in contrast to other approaches to LV. The result indicated, nonetheless, at the same time that the foam can mildly modify the electromagnetic cascades under certain conditions, by suppressing pair creation on nuclei by primary photons. In this addendum, we consider how this modification affects the detection of extensive air shower~(EAS) initiated by an ultrahigh-energy cosmic-ray particle~(viz., a primary hadron), like proton with $E \sim 10^{19}$~eV, given that secondary photon subshowers following $\pi^{0}$ decays could be similarly influenced. We argue that fewer electrons would reach the detector and hence the energy of the primary particle may be underestimated due to foam effects, enhancing in such a way the muon content in EASs. This opens up the possibility of interpreting the alleged ``excess'' of muons, as reported by Auger and Telescope Array collaborations recently and many other experiments on high-energy cosmic rays, with a quantum-gravitational effect. Future observations are anticipated to confirm whether this anomaly really exists.
\end{abstract}

\end{frontmatter}

Ultrahigh-energy cosmic rays~(UHECRs) constitute the most energetic particles known so far, reaching energies of several tens of EeV. After their collisions with atmospheric atoms, a huge cascade of particles or shower with lower energies is generated. Among many particulars its study and understanding reveal, it also offers an energetic window ten times higher than collider ones for new physics. Especially, it may be used to probe the existence of Lorentz-invariance violation~(LV) for the photon, such as might be induced by quantum gravity~(QG) effect from space-time foam: see~\cite{Li:2025wqc} and references therein.

As is well known, a number of observatories~\cite{HiRes:1999ioa,Glushkov:2007gd,PierreAuger:2016nfk,Bogdanov:2018sfw,TelescopeArray:2018eph,Gesualdi:2020ttc,Kalmykov:2022grb,KM3NeT:2024buf} has reported a significant, yet unexplained, discrepancy between the observed muon content in extensive air showers~(EASs) with that predicted by state-of-the-art interaction models. In~\cite{PierreAuger:2016nfk}, the Pierre Auger collaboration~\cite{PierreAuger:2015eyc} reported 30--60\% more muons than expected, for the hadronic component of showers of primary energy, $10^{9.8}<E/\textrm{GeV}<10^{10.2}$. Since this publication, together in particular with the later result of Telescope Array~\cite{TelescopeArray:2018eph}, the problem of the muon ``excess'' in EASs, or, ``muon puzzle''~\cite{Albrecht:2021cxw}, has begun to attract greater attention, suggesting a need for refinements in our current understanding of fundamental~(cosmic-ray) physics. While some experiments noted on the contrary via their measurements~\cite{Fomin:2016kul,KASCADE-Grande:2017wfe,IceCube:2022tla} of the average muon density at 1000~m from shower core $\langle N_{\mu}\rangle$, the recent meta-analysis~\cite{ArteagaVelazquez:2023fda} of Working Group on Hadronic Interactions and Shower Physics again revealed the existence of a significant deficit of muon number $N_{\mu}$ in Monte Carlo~(MC) simulations of air showers~(as compared to the observational data).

These observations provide a basis for discussing a possible~(QG) interpretation of muon excess in UHE air showers that we consider in this \emph{addendum} to~\cite{Li:2025wqc} providing pointers for future refined studies.

Potential modifications of hadronic interactions could be one of the most conservative resolutions to this discrepancy however a description consistent with collider results is still missing. Recently, there has been an increasing interest in explaining the muon puzzle using new-physics models with LVs, that leave the hadronic interactions intact but modify the electromagnetic~(EM) showers of CR EASs~\cite{Klinkhamer:2016evw,Tomar:2017mgc,Duenkel:2023nlk,Diaz:2016dpk,Martynenko:2024rhj}. In particular, such modifications may occur in a string model for space-time foam from QG-induced LV effects on photon waves, as suggested in ref.~\cite{Li:2025wqc}, for the case of showers initiated by a primary \emph{photon} entering the Earth's atmosphere. In this comment, we argue that, for cosmic-ray-~(primary-\emph{hadron}-)produced EASs, the simulated muon content may need to be reevaluated, provided the latter relies upon the primary particle energy reconstructed from the secondary photon-induced EM cascades~(cf. subshowers) which could be influenced by effects of space-time~(D-particle) foam defects in the spirit of~\cite{Li:2025wqc}.

We should mention that the idea of solving muon deficit with LV was first proposed in~\cite{Tomar:2017mgc} in a \emph{superluminal} photon LV scenario, in which case the decay $\pi^{0}\rightarrow\g\g$ is suppressed for neutral pions in the EAS, reducing the average fraction $f_{\textrm{EM}}$ in EM particles per generation. A larger fraction~($f_{h}$) of the primary energy is then released in a hadronic shower where muons are eventually produced, as evidenced from\footnote{It was noted~\cite{Allen:2013hfa} that reducing $f_{\textrm{EM}}$~($\pi^{0}$ energy fraction) through, e.g., pion stability~(or nondecaying $\pi^{0}$'s, as initially considered in~\cite{Antonov:2001xh} but in the frame of Coleman--Glashow LV scheme~\cite{Coleman:1998ti}; see~\cite{Amelino-Camelia:2001com}, for a discussion in a different context), is the only viable option for increasing the predicted muon number without conflicting with observations of the shower maximal depth $X_{\max}$.}
\begin{equation}
\label{eq:1}
f_{h}\sim (1-f_{\textrm{EM}})^{n_{\textrm{gen}}},
\end{equation}
where $n_{\textrm{gen}}$ is the number of generations required for most pions to have energies $\lesssim 100$~GeV~(below which $\pi^{\pm}$'s decay rather than interact)~\cite{Matthews:2005sd}. However, a superluminal photon may decay into an $ee^{+}$ pair modifying the depth of shower maximum $X_{\max}$, and limiting the LV effects~\cite{Diaz:2016dpk}. The lack of modified $\pi^{0}$ decays and vacuum \v Cerenkov effects of electrons inferred from EAS simulations also tightly constrains other field-theory models for LV~\cite{Klinkhamer:2016evw,Duenkel:2023nlk}. Apart from these possibilities, however, a viable way of describing the origin of EAS muon problem, consistent with other results, might be provided by the stringy \emph{D-foam} model of~\cite{Li:2025wqc}, as we now discuss. Main differences from the previous approaches are highlighted.

The potential Lorentz violation is inspired by analysts of quantum space-time/QG~\cite{Li:2025stf}, predicting rich astroparticle phenomenology associated with either the induced energy-dependent speed variation $v(E)$ for high-energy relativistic probes~(say, energetic photons), or ``anomalies'' in particle-interaction processes as a result of LV-deformed dispersion of the particle of energy, $E$, or frequency $\omega=E$ in case of a photon:
\begin{equation}
\label{eq:2}
c^{2}k^{2}-\omega^{2}\simeq s_{n}\omega^{2}\Bigl(\frac{\omega}{E_{\textrm{LV}n}}\Bigr)^{n},
\end{equation}
where $E_{\textrm{LV}n}$ is the $n$th-order LV scale, approaching Planck scale $\Ep\sim 10^{19}$~GeV, while $s_{n}=+1$ or $-1$ for subluminal or superluminal propagation and low-energy probes travel at the constant light speed \emph{in vacuo} $c$~(that we set to unity below). It was suggested~\cite{Amelino-Camelia:1997ieq} that the existence of such LV could best be probed with energetic photons from transient astrophysical sources such as gamma-ray bursters~(GRBs) or active galactic nuclei. Recent researches have suggested that there is a \emph{light-speed variation}, $v=1-\omega/\Elv$, with $\Elv\simeq 3\times 10^{17}$~GeV, $s=+1$~(cf. \emph{subluminal} scenario) for these photons~\cite{Xu:2016zxi,Amelino-Camelia:2016ohi,Liu:2018qrg,Li:2020uef,Song:2024and,Song:2025myx,Song:2025ksi}~(for a review see~\cite{Li:2025stf}), corresponding to LV at $\O(k/\Ep)$, i.e., in Eq.~(\ref{eq:2}) $n=1$, in which case we discard the suffix $n$. However this dispersion effect, if real, would also cause suppression of photon pair creation in the nuclear field~(which is of Bethe--Heitler process), assuming that energy--momentum conservation holds. This modifies cosmic photon showers with changes in conflict with experiment~\cite{Vankov:2002gt,Rubtsov:2016bea,Satunin:2023yvj}, and strict limits have been established, e.g., $\Elv\lvert_{\g\textrm{EAS}}\ \gtrsim 10^{20}$~GeV~\cite{Satunin:2023yvj} with EASs induced by primary $\g$-rays up to $\sim 1$~PeV.

In our recent paper~\cite{Li:2025wqc}, we suggested a D-brane model of space-time foam as a candidate theory that can accommodate the phenomenological finding of light-speed variation without inducing drastic changes to photon showers in air, unlike other~(field-theoretic~\cite{Myers:2003fd}) LV models with subluminal photons~\cite{Vankov:2002gt}. We recall that the presence of D-particle defects in the space-time background of the model induces a deviation of photon dispersion relations from that of relativity~($\omega=k$), that is of a LV energy defect $\d\omega\coloneqq\omega-k$: $\d\omega\lvert_{\textrm{D-foam}}\simeq -\sigma^{2}k^{2}/(2\Md)$, suppressed by a single power of string mass scale, $\Ms\eqqcolon\gs\Md$, where $\gs<1$ is the string coupling assumed to be weak, in which case we expect the D-particle mass $\Md\leq\O(\Ep)\sim 10^{18}$~GeV, and the light-speed variation is interpreted via such a deviation $\d\omega$ with $\sigma^{2}\sim\O(1)$~\cite{Li:2025wqc}. Since the model also predicts \emph{energy fluctuations} during particle \emph{reactions}~\cite{Ellis:2000sf}, modifications of $\g$-ray showers are determined by an interplay between dispersion effects and this latter phenomenon, which is controlled by another factor $\varsigma_{I}$: while a foam with $\sigma^{2}>\varsigma_{I}/2$ would suppress the formation of showers there are no such anomalies if $\varsigma_{I}=2\sigma^{2}$. Bounds for $\Elv$, due to the lack of anomaly in EAS of PeV $\g$-rays~\cite{Satunin:2023yvj} are then evaded, or, at most imply an \emph{upper bound} on the difference,
\begin{subequations}\label{eq:3}
\begin{align}
\label{eq:3a}
(2\sigma^{2}-\varsigma_{I})&\lesssim\frac{10\Ms}{\gs}(\Elv\lvert_{\g\textrm{EAS}})^{-1}\sim 10^{-2},\\
\label{eq:3b}
\Elv\lvert_{\g\textrm{EAS}}&>2.1\times 10^{20}~\textrm{GeV},
\end{align}
\end{subequations}
but these do \emph{not} constrain foam dispersion factor, $\sigma^{2}$. The latter is related to the suppression scale $\Elv$ of light-speed variation from independent~(time-of-flight) studies~\cite{Xu:2016zxi,Amelino-Camelia:2016ohi,Liu:2018qrg,Li:2020uef,Song:2024and,Song:2025myx,Song:2025ksi} of cosmic photons, thereby surviving the shower formation bounds~(\ref{eq:3}). These points have been elaborated in Sec.~2 of ref.~\cite{Li:2025wqc} which form the basis of the discussion in this addendum, so the reader is referred to this section, in particular, for further details.

Let us now consider EASs induced by UHECR primaries of energies $\gtrsim 10^{9.8}$~GeV such as protons with $E\sim 10^{19}$~eV. The first interaction occurs almost immediately as the particles enter the atmosphere, generating secondary products which are intrinsically hadrons, primarily charged and neutral pions. These charged pions subsequently interact with air~\cite{Matthews:2005sd}, while the neutral $\pi^{0}$'s decay promptly to two photons after their production, with lifetimes of~$\O(10^{-17})$~s in the Lorentz-symmetric standard model. For a specific type of LV~\cite{Tomar:2017mgc}, mentioned previously, which predicts the decay suppression of $\pi^{0}$, it is possible that the fraction of the primary energy transferred to decaying $\pi^{0}$'s, and hence to the ensuing EM subshowers is suppressed to yield a desired enhancement of the muon content in cosmic-ray atmospheric cascades in light of~(\ref{eq:1}).

We want to address at this point the problem of whether the same argument can be applied in D-foam models. First we recall that in such models~\cite{Ellis:1999uh,Ellis:2000sf,Ellis:2003sd,Ellis:2004ay,Mavromatos:2005bu,Ellis:2008gg,Li:2021gah,Li:2025wqc}, LV acts as a back-reaction effect of the recoiling D-branes on space-time~(the resulting energy-dependent modification of the metric, and in turn, of the usual dispersion relations~(\ref{eq:2}) with $n=1$, is of Finsler type; $\omega=k+\d\omega$ for photons), hence, it can act only on real particles. But in order for neutral $\pi^{0}$'s created in the earliest part of the EAS formation to undergo radiative decays they must excite modes of the electromagnetic vacuum,\footnote{We note that while the atmosphere is a medium in the rest frame of high-energy pions, $\pi^{0}$ decays in air shower simulations are treated as if they occur in a vacuum~\cite{Allen:2013hfa}.} i.e., virtual photons, which then are not affected by LV. In fact, as a mere process of photon production, $\pi^{0}$ decay is simply \emph{not} influenced by the presence of the foam, according to a ``theorem'' in~\cite{Li:2025wqc}, which explains unmodified bremsstrahlung following pair creation of incoming cosmic $\g$-rays. For specifically stringy reasons we outlined therein, associated with the charge conservation, to hadrons like pions, which are not structureless and whose constituents are charged, D-particle foam actually looks transparent.~(This also allows the theory to escape the very strong constraints on pion LV effect~\cite{PierreAuger:2021mve} established with the muon measurements of Auger
Observatory.)

Since $\pi^{0}$'s do decay immediately, without suppression in our model, they feed the electromagnetic component of the shower in the usual way. However, as it becomes clear from the above discussion, and that of ref.~\cite{Li:2025wqc}, since the ensuing EM subshowers are produced by their decay photons, they could be affected if there were a foam with $\sigma^{2}\neq\frac{1}{2}\varsigma_{I}$. Note that the EAS lepton content at ground depends heavily on these secondary photon subshowers and that, the energies of the photons are $\sim 10^{17}$~eV, 2 orders of magnitude higher than the energy of primary photons ever detected~($\sim$ a few PeV), hence, the corresponding generic LV~(\ref{eq:2})~(or a stringy foam with energy nonconservation) motivated anomaly for $\g$-ray air showers is still not experimentally tested. In case of foam with parameters $\varsigma_{I}<2\sigma^{2}$, the cross-section of pair creation or Bethe--Heitler~(BH) process $\g Z\rightarrow Zee^{+}$ in the Coulomb field of a nucleus $Z$~(the most probable channel of $\g$-ray's first reaction in air) would be suppressed according to Eq.~(14) in~\cite{Li:2025wqc} due to reduced formation length $\lambda_{f}$ of the process. This increases the chance of photons penetrating more material than conventional, corresponding to a larger photon mean free path $\tilde{l}_{\g}$ during EAS evolution:
\begin{equation}
\label{eq:4}
\tilde{l}_{\g}\propto\tilde{\sigma}^{-1}\sim\Bigl(\frac{\lambda_{f}}{\lambda_{f0}}\sigma_{\textrm{BH}}\Bigr)^{-1}>l_{\g},
\end{equation}
where $\tilde{\sigma}$ is the foam modified interaction cross-section; the related $\lambda_{f}$ is shorter than $\lambda_{f0}$ in the Lorentz invariant case. The most important
implication in the EAS context is that since fewer $ee^{+}$ pairs are produced in these deeper showers the average number of electrons registered by the detector, $\langle \tilde{N}_{e}\rangle$, is lower in our case, compared to the standard, foam-free expectation $\langle N_{e}\rangle$.

From these considerations, a crucial observation for solving the muon puzzle arises at this juncture: since the foam induces $\langle\tilde{N}_{e}\rangle<\langle N_{e}\rangle$, the energy of the primary cosmic ray, reconstructed under Lorentz invariance, is actually smaller than its true energy, $E$. To see it more clearly, we first note that hybrid observatories including Auger instruments~\cite{PierreAuger:2015eyc} and the Telescope Array~\cite{TelescopeArray:2018eph} calibrate their energy measurements with fluorescence detectors, and the air fluorescence is determined via shower electrons. The primary energy reconstruction thereupon depends more or less on measuring $N_{e}$~\cite{Martynenko:2024rhj}. We may assume that reconstructed energy $E_{\textrm{rec}}$ of a hadronic EAS primary follows a power-law relation with the registered $\langle N_{e}\rangle_{\textrm{reg}}\propto E_{\textrm{rec}}^{\alpha_{e}}$:
\begin{equation}
\label{eq:5}
\ln\Bigl[\frac{E_{\textrm{rec}}}{\textrm{GeV}}\Bigr]=\alpha_{e}^{-1}\ln\langle N_{e}\rangle_{\textrm{reg}}+\ldots,
\end{equation}
where the $\ldots$ on the r.h.s.\! stand for contributions that are not supposed to be affected other than the $\langle N_{e}\rangle$ term, such as the logarithm of the reconstructed mass number $\ln A_{\textrm{rec}}$ in any actual experimental analysis. Moreover, our stringy model has \emph{no} impact on the number of muons as the latter are born predominantly from $\pi^{\pm}$ decays~\cite{Matthews:2005sd} which are also not altered by D-foam for reasons explained above. We can plausibly use an analogous scaling for mean $N_{\mu}$ as follows: $\langle N_{\mu}\rangle\propto E^{\alpha_{\mu}}$, viz.,
\begin{equation}
\label{eq:6}
\ln\langle N_{\mu}\rangle=\alpha_{\mu}\ln\Bigl[\frac{E}{\textrm{GeV}}\Bigr]+\ldots,
\end{equation}
which produces the simulated values for $\langle N_{\mu}\rangle$ in air shower MCs. $\alpha_{\mu}$ and the power $\alpha_{e}$ in Eq.~(\ref{eq:5}) can be determined by a simulation in the conventional case, utilizing for example CORSIKA~\cite{Heck:1998vt} with EPOS~1.99 and EGS4 as hadronic and EM event generators respectively. However due to the QG-motivated suppression of $\langle N_{e}\rangle$ in EASs as discussed above, it is the reduced $\langle\tilde{N}_{e}\rangle$ that would be identified with the observed $\langle N_{e}\rangle_{\textrm{reg}}$, which leads to a bias in energy reconstruction $E_{\textrm{rec}}<E$. Without incorporating this suppression~(\ref{eq:4}) into the Lorentz-invariant EGS4 model of EM interactions, the expected average muon number, $\langle N_{\mu}\rangle_{\textrm{exp}}$~(cf.~(\ref{eq:6})), from MCs based on the \emph{reconstructed} energy, $E_{\textrm{rec}}$, therefore underestimates the actual EAS muon density in the presence of foam, $\langle N_{\mu}\rangle_{\textrm{exp}}<\langle N_{\mu}\rangle$, as is indeed the case in measurements of the muon signal.

We see that contrary to other proposals to fit Auger and other data, presented earlier in~\cite{Allen:2013hfa,Tomar:2017mgc}, where the increase in $N_{\mu}$ is achieved by modifying $\pi^{0}$ energy fraction~(encoded in EPOS~1.99), our approach does not change the number of muons, thereby avoiding changing $X_{\max}$ which is tightly bounded. Indeed, $\langle N_{\mu}\rangle$ remains the same in both Lorentz-invariant and foam cases for the \emph{true} primary energy, $E$, of the cosmic-ray particle. The muon deficit in simulations of showers simply arises from the biased $\langle N_{e}\rangle_{\textrm{reg}}$-based energy reconstruction due to a ``lack of electrons'' produced in the EM subshower which is suppressed in the case of $\sigma^{2}>\frac{1}{2}\varsigma_{I}$. As noted elsewhere~\cite{Li:2025stf}, on the other hand, anomalous LV $\g$-decays or $e^{\mp}$ \v Cerenkov effects \emph{in vacuo} are in general not allowed by D-particle foam, so the model escapes the strict constraints due to the absence~\cite{Diaz:2016dpk,Duenkel:2023nlk} of these channels in a hadronic EAS.

Certainly, to accurately test the above modifications and to uncover signatures of~(stringy) QG, a full numerical MC shower simulation involving foam-suppressed BH processes of secondary photons should be used. Going into details of such an analysis is far beyond the scope of this paper since our aim here is to offer some crude and heuristic arguments which by themselves establish nothing but have the potential to offer guidance for future attempts of rendering more reliable the proposed solution of the deficit with a theory of stringy foam~\cite{Li:2025wqc}. In fact, our mechanism works in a similar manner to the one~\cite{Martynenko:2024rhj} for a model entailing subluminal LV at $\O(k^{2}/\Ep^{2})$ for photons, see Eq.~(\ref{eq:2}), $\omega^{2}\simeq k^{2}+2k\d\omega\vert_{\textrm{LV}2}$ with $\d\omega\vert_{\textrm{LV}2}=-[\omega^{3}/(2E_{\textrm{LV}2}^{2})]$. In ref.~\cite{Martynenko:2024rhj}, the corresponding modifications of the EM subshowers were implemented by suppressing the secondary particles of EGS4, and it has been shown that a LV scale of $E_{\textrm{LV}2}\sim 10^{16}$~GeV is required to solve the muon puzzle, assuming a perfect primary composition reconstruction. Moreover, the nonobservation of a much stronger discrepancy between EAS data and simulations has been used to constrain such $n=2$ LVs in photon dispersion $E_{\textrm{LV}2}>2.4\times 10^{14}$~GeV based on meta-analyses of~(Auger) data~\cite{ArteagaVelazquez:2023fda}.

Note that for those general and/or field-theoretic scenarios for LV photons~\cite{Vankov:2002gt,Myers:2003fd,Rubtsov:2016bea,Satunin:2023yvj,Martynenko:2024rhj} where there is no mechanism for energy nonconservation~(unlike our foam situation) despite the presence of modified dispersions~(\ref{eq:2}), $\g$-ray shower suppression effect occurs as a direct consequence of dispersion modifications $\d\omega\vert_{\textrm{LV}n}$ of the photon. In order to obtain a first order-of-magnitude constraint in the linear subluminal LV case $n=1$, which is largely favored by GRB data in light of the indication of light-speed variation~\cite{Xu:2016zxi,Amelino-Camelia:2016ohi,Liu:2018qrg,Li:2020uef,Song:2024and,Song:2025myx,Song:2025ksi} mentioned above, we assume that a detailed numerical analysis could establish a bound in $\d\omega\lvert_{\textrm{LV}}$ as ref.~\cite{Martynenko:2024rhj} established on $\d\omega\lvert_{\textrm{LV2}}$, and the following comparison may provide a crude exploratory estimate:
\begin{equation}
\label{eq:7}
\d\omega\vert_{\textrm{LV}}=-\Bigl(\frac{\omega^{2}}{2\Elv}\Bigr)\simeq\d\omega\vert_{\textrm{LV}2}=-\Bigl(\frac{\omega^{3}}{2E_{\textrm{LV}2}^{2}}\Bigr).
\end{equation}
Accordingly, we estimate $\Elv\gtrsim E_{\textrm{LV}2}^{2}/\omega$, where we plug in $\omega\simeq 10^{17}$~eV, and, using the fact that Auger data excludes $E_{\textrm{LV}2}\leq 2.4\times 10^{14}$~GeV, obtain,
\begin{equation}
\label{eq:8}
\Elv\gtrsim 6\times 10^{20}~\textrm{GeV}.
\end{equation}
This exceeds the limit~(\ref{eq:3b}) obtained from the detections of normal EASs of PeV photons by a factor of $\sim 3$, and also, is about 1 or 2 orders of magnitude greater than the natural order $\O(\Ep)$ of the scale. We emphasize that although this may be commonly used as an indication that linear LV frameworks entailing subluminal photons are ruled out on naturalness grounds, this is not the case for the effect arising in our D-brane model. Due to energy fluctuations~(proportional to $\varsigma_{I}$) caused by the foam, \emph{no} such constraint can be derived for $\d\omega\vert_{\textrm{D-foam}}$, and as becomes clear from~\cite{Li:2025wqc} and the earlier summary, the bound~(\ref{eq:8}) is no longer imposed on the LV scale in~(\ref{eq:2}) that corresponds to $\sigma^{2}$, but on $2\sigma^{2}-\varsigma_{I}$ in our case, as evidenced from Eq.~(\ref{eq:3a}).

It is clear from this argument that in such a string model for space-time foam that entails LV in photon propagation, there would be compatibility with the aforementioned hint of light-speed variation from independent studies of cosmic photon time-of-flights which require a subluminal LV scale ($\sim10^{17}$~GeV) several orders of magnitude lower than that set by considerations on EAS muon content. On the other hand, using the best-fit value, $E_{\textrm{LV}2}=1.9\times 10^{16}$~GeV~(for the quadratic case~\cite{Martynenko:2024rhj}) capable of fitting the relevant measurements of mean $N_{\mu}$, which would correspond to a linear suppression scale of $\sim 10^{24}$~GeV~(cf.~(\ref{eq:7})), we are now able to make an analogous estimate for the possible values that the foam parameters might have to take in order to explain the muon deficit:
\begin{equation}
\label{eq:9}
2\sigma^{2}-\varsigma_{I}\sim 10^{-6}.
\end{equation}
It does not impose any constraint \emph{per se} but only indicates, that~(for traditionally Planckian D-particle mass scales assumed earlier) $\varsigma_{I}$, like $\sigma^{2}$, the parameter governing photon propagation, is of $\O(1)$; this can be easily satisfied, since in such string models, both parameters of the foam are freely adjustable~\cite{Li:2025wqc}. Importantly, Eq.~(\ref{eq:9}) also respects the upper bound~(\ref{eq:3a}), implying that the QG effect introduced here is weak enough to account for muon puzzle while avoiding inducing anomalies in EASs initiated by primary photons of PeV energies~(the highest-energy ones currently detected), in agreement with their observations.

We close by noting that the quantum-gravitational effect from a string D-foam model -- compatible with the Lorentz-violating speed variation of cosmic photons -- could provide a solution for the puzzling discrepancy of the measured and simulated muon component of EASs. In this model, the $\g$--air first reaction is modified in a phenomenologically viable way, implying a so-induced modification to primary energy reconstruction used in simulations of hadron-induced EAS. We emphasize, however, that this proposal, along with our estimate~(\ref{eq:9}), cannot be regarded as definitive at present. A careful assessment is required, involving detailed numerical comparisons between data and MC predictions, with foam effects explicitly incorporated into simulation codes. There is no reason to jump to any conclusions, also because those existing claims regarding the muon puzzle remain far from conclusive. We await future cosmic-ray observations which may clarify whether such discrepancies indeed manifest in UHE air showers.

\section*{Acknowledgments}

This work is supported by Beijing Natural Science Foundation under Grant No.~1264066. I acknowledge the China Postdoctoral Science Foundation---CPSF under Grant No. 2024M750046---and a \emph{Boya} Postdoctoral Fellowship of the Peking University for partial support. The author was also funded in part by Postdoctoral Fellowship Program~(Grade B) of CPSF~(Grant No.~GZB20230032) and National Natural Science Foundation of China~(Grant No.~12335006). I am grateful for activities of the European Union COST Action~(CA23130)---Bridging high and low energies in search of quantum gravity~(BridgeQG).

\section*{Note added}

These observations were obtained after we addressed the referee's query during the review of the letter ``Shower formation in the presence of a string-inspired foam in space-time''~\cite{Li:2025wqc}. However, since these remarks concern secondary EM subshower anomaly in EASs, produced by cosmic rays, instead of primary $\g$-rays discussed in the letter, they were not presented therein. Recently, we have noticed the muon excess in cosmic ray data which may relate to this anomaly and, therefore, believe it appropriate and timely to suggest this note as an addendum to our published work given we had soon realized these points.


\end{document}